\input{aipcheck}

\documentclass[
    ,final            
  ]
  {aipproc}

\layoutstyle{6x9}

\usepackage{aas_macros}
\usepackage{amssymb}
\usepackage{txfonts}
\usepackage{natbib}
\usepackage{multicol}

\def\aap{A\&A}
\def\apj{ApJ}
\def\apjl{ApJ}
\def\apjs{ApJS}
\def\aj{AJ}

\begin{document}

\title{The evolution of obscured accretion}

\classification{98.54.Cm}
\keywords      {X-rays: galaxies - galaxies: active - X-rays: diffuse background}

\author{Roberto Gilli}{
  address={INAF - Osservatorio Astronomico di Bologna, Via Ranzani, 1 - 40127, Bologna, Italy}
}
\author{A. Comastri}{
  address={INAF - Osservatorio Astronomico di Bologna, Via Ranzani, 1 - 40127, Bologna, Italy}
}
\author{C. Vignali}{
  address={Universit\`a degli Studi di Bologna, Dip. di Astronomia, Via Ranzani, 1 - 40127, Bologna, Italy}
}
\author{P. Ranalli}{
  address={Universit\`a degli Studi di Bologna, Dip. di Astronomia, Via Ranzani, 1 - 40127, Bologna, Italy}
}
\author{K. Iwasawa}{
  address={INAF - Osservatorio Astronomico di Bologna, Via Ranzani, 1 - 40127, Bologna, Italy}
}

\begin{abstract}


Our current understanding of the evolution of obscured accretion onto supermassive black holes is reviewed. We consider the literature results on the relation between the fraction of moderately obscured, Compton-thin AGN and redshift, and discuss the biases which possibly affect the various measurements. Then, we discuss a number of methods - from ultradeep X-ray observations to the detection  of high-ionization optical emission lines - to select the population of the most heavily obscured, Compton-thick AGN, whose cosmological evolution is basically unknown.
The space density of heavily obscured AGN measured through different techniques is discussed and compared with the predictions by current synthesis models of the X-ray background. Preliminary results from the first half of the 3 Ms XMM observation of the Chandra Deep Field South (CDFS) are also presented. The prospects for population studies of heavily obscured AGN with future planned or proposed X-ray missions are finally discussed.

\end{abstract}

\maketitle


\section{BH/galaxy co-evolution}

The observed scaling relations between the structural properties of massive galaxies - such as their bulge mass, luminosity,
and stellar velocity dispersion - and the mass of the supermassive black holes (SMBHs) sitting at their centers suggest that
the build-up of galaxies and the growth of black holes are closely related, which is often referred to as 
BH/galaxy co-evolution. Since the BH growth is thought to happen primarily through efficient accretion phases accompanied by the release of kinetic and radiative energy, part of which can be deposited into the galaxy interstellar medium, 
Active Galactic Nuclei (AGN) are believed to represent a key phase across a galaxy's lifetime. Support for this hypothesis comes
from several lines of evidence such as i) the match between the mass function of SMBHs grown through AGN phases and the one observed in local galaxies \cite{marconi04,shankar04}; ii) the cosmological ``downsizing'' of both nuclear activity and star formation \cite{ueda03,cowie96};  iii) the detection of large quantities of gas outflowing from active nuclei (see \cite{elvis06} for a recent review).
A number of semi-analytic models to explain the BH/galaxy co-evolution have been proposed over about the past decade \cite{kh00}. 
These models follow the evolution and growth of dark matter structures across cosmic time, either through the Press-Schechter
formalism or through N-body simulations, and use analytic recipes to treat the baryon physics within the dark matter halos.
A common assumption of these models is that mergers between gas-rich galaxies trigger nuclear activity and star formation.
Recently, a BH/galaxy evolutionary sequence associated to ``wet'' galaxy mergers has been proposed \cite{hop08}, in which an initial 
phase of vigorous star formation and obscured, possibly Eddington limited, accretion is followed by a phase
in which the nucleus first gets rid of the obscuring gas shining as an unobscured QSO, then quenches star formation, and eventually 
fades, leaving a passively evolving galaxy. 

\section{The obscured AGN fraction}

The various BH/galaxy co-evolution models have been successful in many respects, being able to explain the local BH/galaxy scaling relations, the local BH mass function, the QSO luminosity function and clustering \cite{hop08,marulli08}. 
Further testing is however needed, given the large numbers of parameters involved in the modeling. A built-in feature of the BH/galaxy evolutionary sequence is that an obscured 
accretion phase preceeds a clean accretion phase, at least in powerful, QSO-like objects. Is it therefore reasonable to expect that the fraction of obscured AGN was higher in the past? This depends on many parameters such as i) the physical scale of the absorbing gas and how it is driven towards the BH; ii) the relative timescales of the obscured and unobscured phases; iii) whether the absorbed-to-unabsorbed AGN transition occurs also in low-mass/low-luminosity objects (i.e. Seyfert galaxies). 

Not many theoretical predictions on the evolution of the obscured AGN fraction are available in the literature.
Depending on the assumptions, some models predict an increase of the obscured AGN fraction with redshift \cite{menci08}, some others do not \cite{lamastra08}.

From an observational point of view the situation is also debated. An increase of the obscured AGN fraction
with redshift among X-ray selected AGN has been for instance observed by \cite{lafranca05,tu06,h08,trump09}, but 
other works did not find any evidence of this trend \cite{ueda03,dp06,gch07}. Much of the uncertainty can be related
to the lack of statistically large samples of AGN with high-quality X-ray spectra. Indeed, using small samples which
cannot be split into narrow luminosity and redshift bins, it is often complicated to disentangle any 
redshift-dependent from luminosity-dependent effects (it is generally agreed that the obscured AGN fraction decreases with luminosity, but the slope and normalization of this relation differ from paper to paper).
Furthermore, the absorption column density is often estimated either from the X-ray hardness ratio, or from the absence of broad emission lines in the optical spectrum. Both methods suffer from caveats and limitations \cite{brusa10}, and ideally the absorbing column should be measured through X-ray spectra with good photon statistics.

\begin{figure}
  \includegraphics[height=.30\textheight]{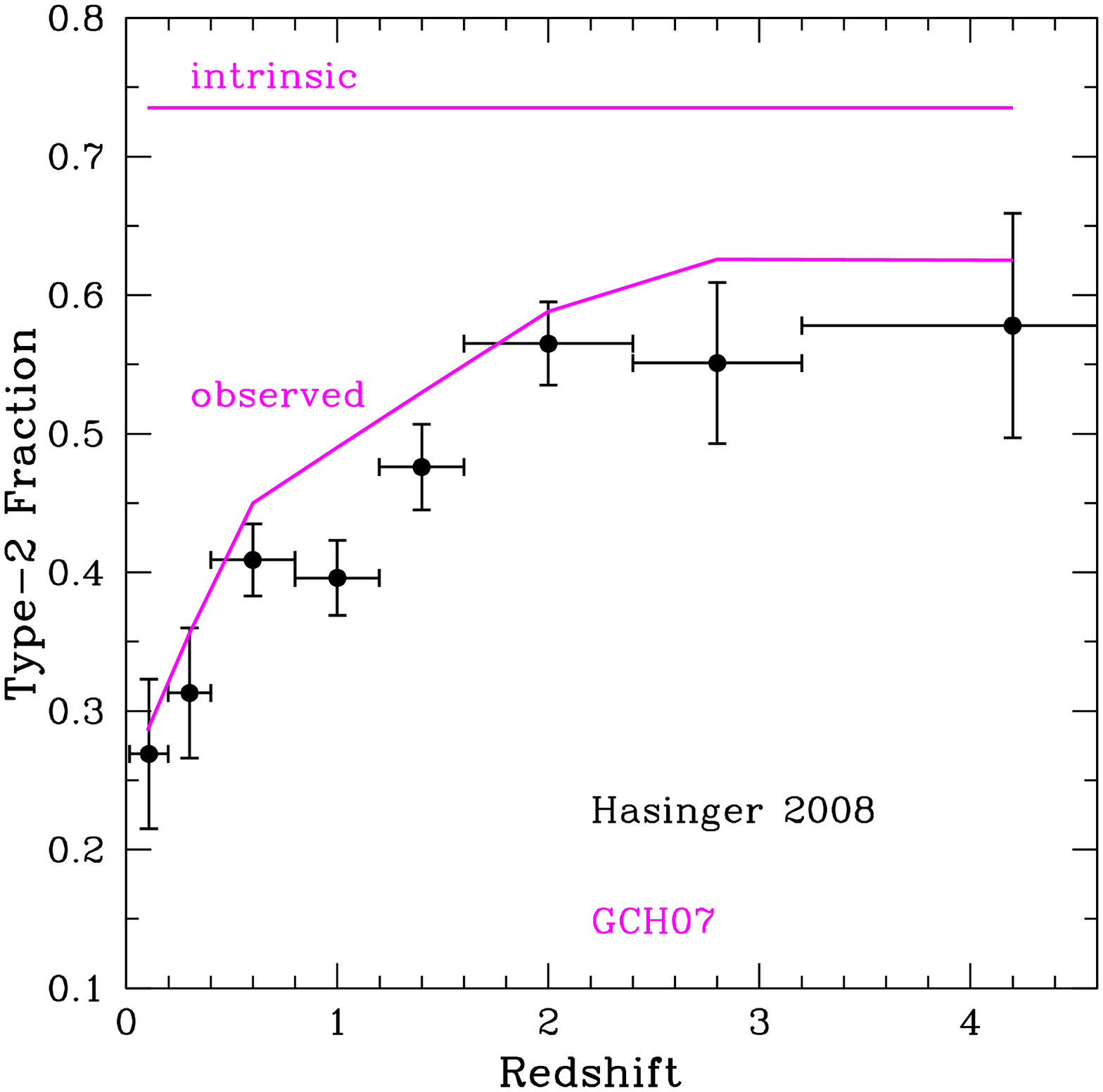}
\includegraphics[height=.30\textheight]{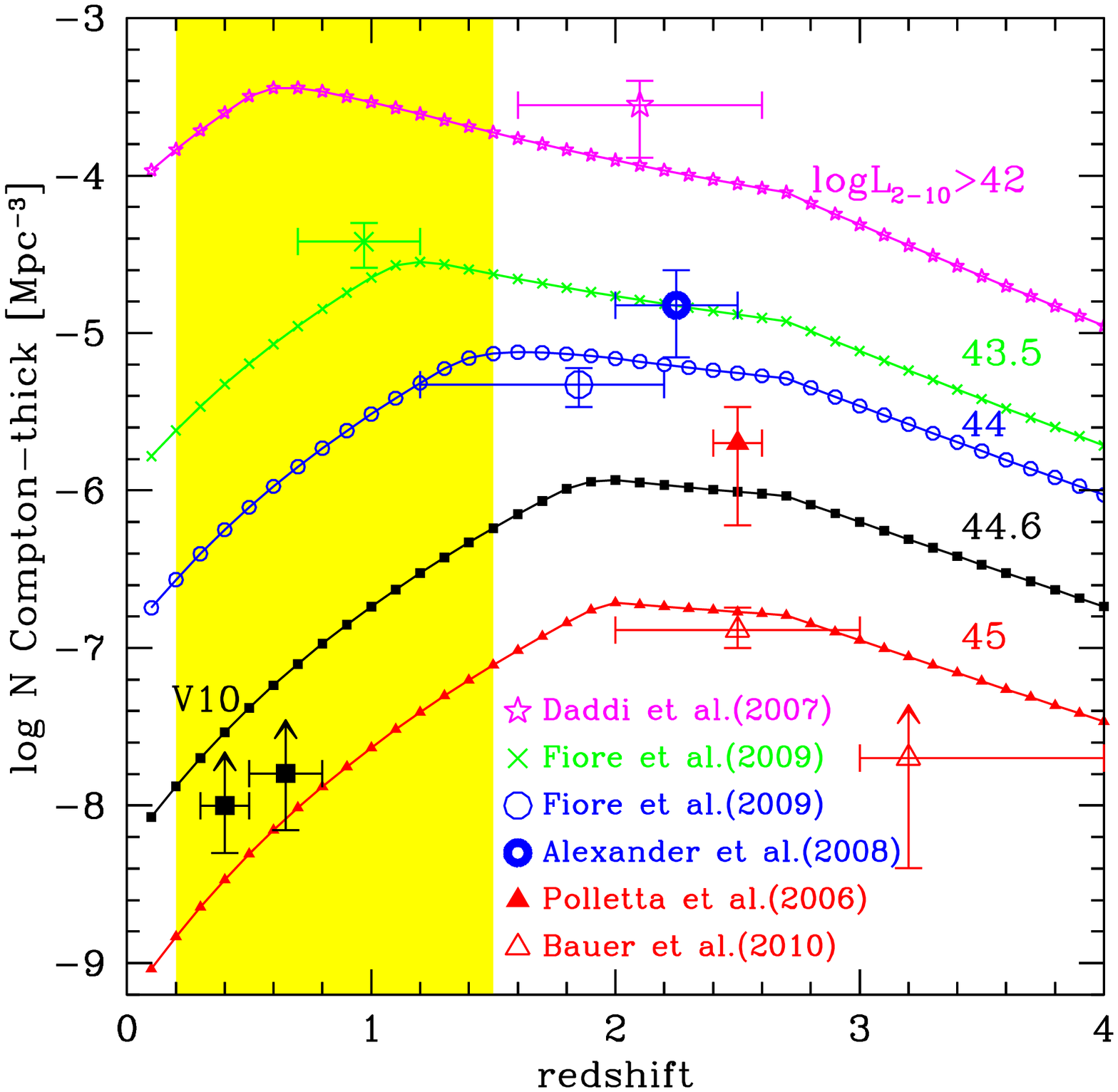}
  \caption{{\it Left panel}: fraction of obscured AGN with $L_{2-10}\sim10^{44}$ erg~s$^{-1}$ as observed by \cite{h08} (Hasinger 2008, datapoints) compared with the model expectations by \cite{gch07} (GCH07, solid lines).  The intrinsic fraction of obscured AGN with $L_{2-10}=10^{44}$ erg~s$^{-1}$,
assumed to be constant with redshift in \cite{gch07}, is shown as an horizontal line. When folding this {\it intrinsic} fraction with the sky coverage
curve of the AGN sample in \cite{h08}, because of selection and K-correction effects, the {\it observed} bent solid curve is obtained. {\it Right panel}: space density of CT AGN as measured through [O~III]5007 selection (datapoints at $z\leq0.7$) and through IR selection ($z\geq0.9$).  Model curves for CT AGN with different X-ray luminosity (as labeled) are from \cite{gch07}: each datapoint should be compared with the curve marked by the same, smaller symbol. The shaded region shows the redshift range accessible to [Ne~V]3426 selection in optical spectroscopic surveys; CT AGN candidates selected through X-ray spectroscopy in the XMM-CDFS cover the same redshift range and further extend to higher redshifts. Adapted from \cite{v10}.} 
\end{figure}

An attempt to measure the evolution with redshift of the obscured fraction of AGN as a function of their X-ray luminosity
has been recently done by Hasinger \cite{h08}, who compiled a sample of $\sim 1300$ AGN selected in the 2-10 keV band and split it into 8 luminosity bins. In Fig.~1 ({\it left}) is shown the evolution of the obscured AGN fraction at luminosities around $L_{2-10}\sim 10^{44}$ erg~s$^{-1}$ as measured in \cite{h08}. Although some approximated
correction has been performed (see Section 6.2 in \cite{h08}), this is essentially an {\it observed} fraction which still suffers from selection biases against obscured objects: because of K-correction effects, this bias is less severe towards higher redshifts, when the photoelectric cut-off moves towards lower frequencies. We verified the strength of
K-correction effects by considering the X-ray background (XRB) model by \cite{gch07}, which matches the obscured AGN fraction at log$L_{2-10}=44$, and in which this fraction is constant with redshift. By folding the logN-logS relations
of obscured and unobscured AGN with log$L_{2-10}=44$ with the sky coverage curve of \cite{h08} we found the trend shown in Fig.~1 ({\it left}): an {\it intrinsic} obscured fraction which is constant with redshift translates into an {\it observed} fraction which is increasing with redshift. Thus, the trend at log$L_{2-10}=44$ found in \cite{h08} might be mostly driven by K-correction 
effects rather than by a truly evolving obscured AGN fraction. The same conclusion also holds at lower luminosities. A possible way to reconcile this result with the redshift evolution found in \cite{lafranca05} and \cite{tu06}, who take into account selection effects in their
estimates but do not split their samples into narrow luminosity bins, is to hypothesize that the obscured fraction increases with redshift {\bf only} for very luminous QSOs. Indeed, the luminosity dependence of the obscured AGN fraction in 
\cite{gch07} appears to be too shallow when compared with recent results \cite{h08,brusa10}, overestimating the number of obscured QSOs with log$L_x\sim45$: a too large number of local obscured QSOs assumed in \cite{gch07} might mask
any increase with redshift. The possibility of a stronger evolution of the obscured AGN fraction for more luminous sources has, for instance, been proposed in the blast-wave semi-analytic model by \cite{menci08}.

\section{The most heavily obscured AGN}

If the overall evolution of obscured AGN is uncertain, the evolution of the most heavily obscured and elusive ones, the so-called Compton-Thick AGN ($N_H>10^{24}$cm$^{-2}$; hereafter CT AGN), is completely unknown. Only $\sim$50 {\it bona-fide} CT AGN are known in the local Universe \cite{c04,rdc08}, but their abundance has nonetheless been estimated to be comparable to that of less obscured ones \cite{guido99}. 
Also, synthesis models of the XRB predict that from $\sim$10 to $\sim$30\% of the XRB peak emission at 30 keV is due to the emission of CT AGN integrated over all redshifts. In the absence of any information, the luminosity function and evolution of CT AGN in XRB synthesis models have been usually assumed to be equal to those of less obscured ones.

In recent years there have been many attempts to constrain the space density of CT AGN in different luminosity and redshift intervals exploiting different selection techniques.
Very hard ($>10$ keV) X-ray surveys are still limited in sensitivity and are just sampling the local Universe \cite{malizia09}. As a result, the population of local CT AGN detected by INTEGRAL and Swift is producing only a tiny fraction ($<1\%$) of the XRB.
To sample AGN at higher redshifts, and in particular at $z\sim 1$, where the bulk of the ``missing'' XRB is thought to be produced, one needs to rely on deep X-ray surveys in the 2-10 keV band and try to select CT AGN either directly from X-ray spectroscopy, or by comparing the measured, obscured X-ray emission (if any) with some other indicator of the
intrinsic nuclear power: IR selection can track the nuclear emission as reprocessed by the dusty absorber; high-ionization, narrow optical emission lines can sample scales free from the nuclear obscuration. X-ray stacking of IR-selected, but X-ray undetected, sources has been used to estimate the space density of CT AGN at $z \sim1-2$ \cite{daddi07,alex08,fiore09,trei09_ct}. The comparison between the [O~III]5007 and X-ray flux 
has been used to select X-ray underluminous QSOs and then estimate the density of CT QSOs at $z=0.5$ \cite{v10}. 
A summary of these and other measurements, compared with the predicted space densities of CT AGN by \cite{gch07}, is
shown in Fig.~1 ({\it right}).  To sample the population of CT around $z\sim 1$, in \cite{neon} we recently devised a selection method based on the [Ne~V]3426 emission line, which can be applied to optical spectroscopic surveys with deep X-ray coverage to search for CT objects at $z>0.8$, i.e. at redshifts not reachable with [O~III]5007 selection. We calibrated an X-ray to  [Ne~V]  flux diagnostic ratio (X/NeV) based on a sample of $\sim 70$ local Seyferts, verifying that objects with X/NeV$<$15 are almost invariably Compton-Thick. We then applied this diagnostic to several samples of obscured and unobscured QSOs: for the objects at $z<0.8$ presented in \cite{v10} we found an excellent agreement between the absorption estimated from the X/OIII and the X/NeV ratio. Furthermore, despite the very low statistics, $\approx 50\%$ of the [Ne V]-selected type-2 QSOs at $z>0.8$ appear to be good CT candidates.

[Ne~V], [O~III], and IR selection are all indirect ways to select CT AGN, since the CT nature of an object is inferred from the faintness of its X-ray emission relative to an indicator of the intrinsic power, and then suffer from a number of systematics like reddening in optically selected sources and non-simultaneous data acquisition.
To sample {\it bona fide} CT AGN over a broad redshift range and, in particular, at $z \sim 1$, very deep X-ray exposures are needed to collect photons and allow X-ray spectral analysis. This is indeed one of the goals of the deep XMM survey in the CDFS (P.I. A. Comastri). The exposure time, cleaned of background flares, currently amounts to $\sim1.8$ Ms and by the end of the program, foreseen by March 2010, should reach $\sim2.5$  Ms. A number of CT candidates, identified in the 1 Ms CDFS catalog
on the bases of their flat (low quality) spectrum \cite{tozzi06}, are being confirmed as such by the higher quality XMM spectra (Comastri et al. in prep.), including  the well-known CT candidate CDFS-202 at z=3.7 \cite{norman02}. Further significant improvements are expected by the forthcoming DDT extension of the Chandra CDFS exposure from 2 Ms to 4 Ms.

\section{Future prospects}

Current deep (Ms) X-ray exposures are limited to very small areas ($<0.4$ deg$^2$), and do not allow construction of sizeable
samples of distant, {\it bona-fide} CT AGN. A number of X-ray missions are being developed or proposed for the near to mid-term future which should address this issue: NuSTAR, ASTRO-H (approved by NASA and JAXA, respectively), EXIST and NHXM (proposed to NASA and ASI, respectively; see these Proceedings)  will primarily focus on the high-energy ($>10$ keV) radiation, which is the least affected by absorption. Their sensitivity is expected to be up to $\sim 2$ orders of magnitude better than current instrumentation. The redshift distribution of CT objects collected by these missions will peak at $z\sim 0.3-0.4$, and will allow population studies of CT objects at redshifts $z<1$. The proposed International X-ray Observatory (IXO, \cite{ixorfi1}), thanks to its large effective area (3~m$^2$ at 1 keV) has the potential to provide very high quality spectra for distant CT AGN, but the small ($<0.1$ deg$^2$) field of view (FOV) might represent a limitation to collect large object samples.  A mission which is certainly expected to allow population studies of distant (up to $z\sim4$) CT AGN, thus providing a measure of their cosmological evolution, is the proposed Wide Field X-ray Telescope \cite{murray08spie}, which has a unique combination of wide FOV (1 deg$^2$), large effective area (1~m$^2$ at 1 keV), sharp angular resolution ($\sim5"-10"$) constant across the FOV, and low instrumental background to provide deep X-ray surveys in the 0.5-7 keV band over large portions of the sky. The equivalent of $\sim$2000 XBootes fields, $\sim3000$ C-COSMOS fields and $\sim$1000 2 Ms CDFS fields is expected to be returned by WFXT over 5 years of operation. As an example of these surveys' potential, based on the current design and observing strategy, the WFXT project is expected to return a sample of $\sim 500$ objects at $z>1$ which are {\it bona fide} CT AGN, i.e. with more than 500 net counts in the 0.5-7 keV band \footnote{For comparision, depending on the observing strategy and background level, in 5 years of operation IXO is expected to return from $\lesssim$100 {\it bona-fide} CT AGN at $z>1$ to a sample as large as the WFXT one.} (the number of simple detections of CT AGN will be obviously much larger). In Fig.~2 we show two such objects (CDF-202 and CDFS-153, another CT AGN, at $z=1.53$, found in the XMM-CDFS)  simulated using the WFXT response matrices for the goal design. It is evident that 500 X-ray photons are sufficient to unambiguously reveal their CT nature. Furthermore, WFXT is expected to return a global sample of 300,000 AGN in the redshift range $z=0-5$ with even larger photon statistics ($>1000$ net counts), for which a robust measurement of the column density is possible. Among the many applications, such an object sample will definitely allow measurement of the evolution of obscured accretion as a function of AGN luminosity.
\begin{figure}
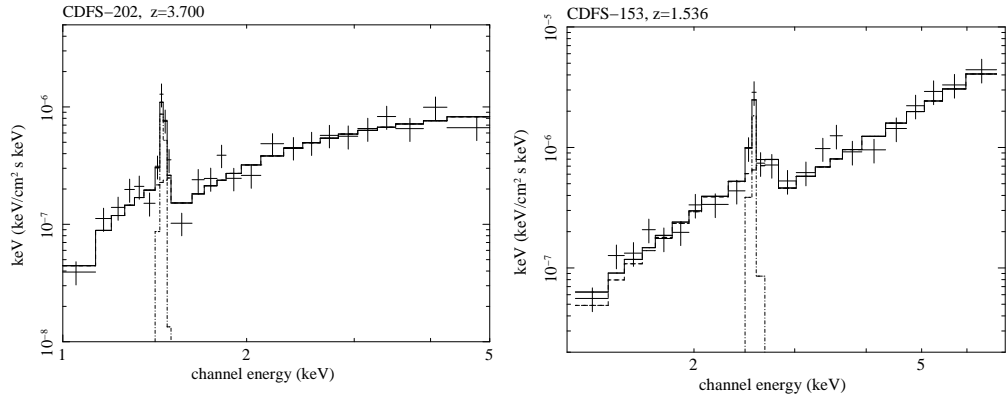

  \includegraphics[height=.3\textheight, angle=270]{gilli_fig2left.ps}
  \includegraphics[height=.3\textheight, angle=270]{gilli_fig2right.ps}
  \caption{400ks WFXT simulated spectra of CDFS-202 and CDFS-153, two high-$z$ CT AGN with $f_{2-10}>10^{-15}$ erg~cm$^{-2}$~s$^{-1}$ observed in the XMM-CDFS. The 100 deg$^2$ WFXT deep survey \cite{murray08spie} is expected to reveal about 500 objects like these - in terms of obscuration and photon statistics - at $z>1$.}
\end{figure}
\begin{theacknowledgments}
We gratefully acknowledge the members of the XMM-CDFS and WFXT 
collaborations.
\end{theacknowledgments}


\bibliographystyle{aipproc}   


\IfFileExists{\jobname.bbl}{}
 {\typeout{}
  \typeout{******************************************}
  \typeout{** Please run "bibtex \jobname" to optain}
  \typeout{** the bibliography and then re-run LaTeX}
  \typeout{** twice to fix the references!}
  \typeout{******************************************}
  \typeout{}
 }

\end{document}